\documentclass[12pt, draftclsnofoot, onecolumn]{IEEEtran}
	\usepackage{pifont,bm,multicol,amsfonts,amsmath,color,amssymb,graphicx, epsfig,cite,psfrag,subfigure,algorithm,enumerate,stfloats,algorithm,algorithmic,epstopdf,balance}
\usepackage{tabularx}
	\usepackage{pifont,bm,multicol,amsfonts,amsmath,color,amssymb,graphicx, epsfig,cite,psfrag,subfigure,algorithm,enumerate,stfloats,algorithm,algorithmic,epstopdf,balance}

\begin{document}
\title{Cascaded Channel Estimation for RIS Assisted mmWave MIMO Transmissions} 	
\author{Yushan Liu, Shun Zhang, {\emph{Senior Member, IEEE,}} Feifei Gao, {\emph{Fellow, IEEE,}} Jie Tang, {\emph{Senior Member, IEEE,}} and Octavia A. Dobre, {\emph{Fellow, IEEE}}

\thanks{Y. Liu and S. Zhang are with the State Key Laboratory of Integrated Services Networks, Xidian University, Xi'an 710071, P. R. China (Email: ysliu\_97@stu.xidian.edu.cn; zhangshunsdu@xidian.edu.cn). F. Gao is with Department of Automation, Tsinghua University, State Key Lab of Intelligent Technologies and Systems, Tsinghua University, State Key for Information Science and Technology (TNList) Beijing 100084, P. R. China (Email: feifeigao@ieee.org). J. Tang is with the School of Electronic and Information Engineering, South China University of Technology, Guangzhou 510641, China (E-mail: eejtang@scut.edu.cn). O. A. Dobre is with Faculty of Engineering and Applied Science, Memorial University, St. John's NL AIC-5S7, Canada (Email: odobre@mun.ca).}
}

\maketitle
\begin{abstract}

Channel estimation is challenging for the reconfigurable intelligence surface (RIS) assisted millimeter wave (mmWave) communications.
Since the number of coefficients of the cascaded channels in such systems is closely dependent on the product of the number of base station antennas and the number of RIS elements, the pilot overhead would be prohibitively high.
In this letter, we propose a cascaded channel estimation framework for an RIS assisted mmWave {multiple-input multiple-output} system, where the wideband effect on transmission model is considered.
Then, we transform the wideband channel estimation into a parameter recovery problem and use a few pilot symbols {to detect the channel parameters} by the Newtonized orthogonal matching pursuit algorithm.
Moreover, the Cramer-Rao lower bound on the channel estimation is introduced.
Numerical results show the effectiveness of the proposed channel estimation scheme.

%As a promising technology, reconfigurable intelligence surface (RIS) can broaden signal coverage, enhance transmission and be applied in the millimeter wave (mmWave) multiple-input multiple-output (MIMO) communication systems.
%In this letter, we solve the problem of wideband channel estimation for an RIS assisted mmWave MIMO system, where the wideband effect is considered.
%Then, the channel estimation problem is expressed as a parameter recovery problem that is tackled by the Newton orthogonal matching pursuit.
%Moreover, the Cramer-Rao lower bound on the channel estimation is introduced.
%Numerical results show the effectiveness of the proposed channel estimation scheme.

\end{abstract}

\maketitle
\thispagestyle{empty}
\vspace{-1mm}


\section{Introduction}

{With wide frequency bands, millimeter wave (mmWave) communications can provide unprecedented gigabits-per-second data rates and satisfy the rapidly growing transmission speed demand of wireless communications \cite{reviewer1_2}.}
However, radio signals in mmWave bands are sensitive to the obstacles and suffer from the severe path loss.
To address this issue,
%massive multiple-input multiple-output (MIMO) technology is applied to combat the path loss, which meanwhile improves the spectral and energy efficiencies \cite{Ma_J_SBL_Time_varing}.
%In recent years, many efforts have been devoted to applying massive MIMO in mmWave communications.
%Since channel state information (CSI) is crucial for massive MIMO systems, various channel estimation techniques \cite{MuyeLi_Time_varing,TDDFDD,time_jianpeng} have been developed for mmWave communications to exploit channel sparsity in angle domain and delay domain.
%In \cite{TDDFDD}, channel estimation is transformed to a sparse signal recovery problem by exploiting sparse scattering property of mmWave channels.
%However, the acquisition of accurate CSI at the base station (BS) would cause tremendous hardware cost and high power consumption in current mmWave massive MIMO systems \cite{otfs_muye,yushanliu_JSAC}.
%In order to break through this bottleneck and achieve more sustainable communications in next-generation mobile networks,
many large-antenna based technologies, such as massive multiple-input multiple-output (MIMO) \cite{jianpengma} and
reconfigurable intelligent surface (RIS) assisted MIMO, have been explored.
Different from radio frequency chain based MIMO systems, RIS is formed as
{an artificial planar structure} with integrated electronic
circuits, and can be programmed to manipulate an incoming
electromagnetic field in a wide variety of functionalities.
It consists of a large number of reconfigurable reflective elements, which
can induce an adjustable independent phase shift on the incident signal \cite{reviewer1_3}.
Recent research results show that RIS assisted MIMO can achieve similar or even higher communication performance gains {with much smaller} hardware cost than massive MIMO.

%The amplitude and phase information at  each RIS element can be adjusted independently, so as to digitally manipulate electromagnetic waves and realize signal propagation direction regulation and in-phase superposition in three-dimensional space, which will improve the quality of received signals and the performance of wireless communication \cite{RIS1}, \cite{RIS2}.
%Hence, incorporating
%RISs in wireless networks has
%been advocated as a revolutionary means to transform
%any naturally passive wireless communication environment, including
%the set of objects between a transmitter and a receiver,  to an active one, and has become a hot research topic.
%
%Among these new technologies, RIS assisted MIMO \cite{shunbo zhang,ODE_CNN} is considered to be very promising to

{Generally}, RIS usually works in the passive state and has no signal processing capability \cite{reviewer1_1}.
Hence, channel estimation in RIS assisted wireless systems is more challenging than that in traditional systems.
In \cite{Matrix,RIS_esti}, they examined the cascaded channel estimation over the RIS aided MIMO systems.
%He {\it{et al.}} realized the channel estimation for the RIS aided mmWave MIMO systems with a two-stage scheme.
In \cite{Hu}, Hu {\it{et al.}} proposed a novel location information aided channel estimation method, which substantially reduced the estimation overhead.
However, in the large-scale antenna systems, different antennas at the same sampling time would receive different time-domain symbols from the same physical path  due to the large propagation delay of electromagnetic waves travelling across the whole antenna array, which is known as the spatial wideband effect \cite{boleiwang}. In such case, the RIS assisted MIMO channel model in the above researches, which only considers phase difference and ignores delay difference among the transmission signals {at different RIS elements and BS antennas, are not applicable anymore.} Moreover, the algorithms based on such models, such as for the channel estimation and parameter recovery, need to be revised.

%Furthermore, the spatial wideband effect due to the large bandwidth considered, the delay difference among the received signals at different antennas in this case cannot be ignored anymore.

In this letter, we investigate the wideband channel estimation scheme for an RIS assisted mmWave MIMO system, which takes the wideband effect into consideration.
The RIS assisted channels  are depicted as the functions of physical parameters, including the angle of arrival/departure (AoA/AoD), the time delay and the complex gain. {The  phase and delay differences} among the received signals at different  BS antennas  and RIS elements are considered.
Then, the frequency response of the received signal is derived
and the channel estimation is formulated as a parameter recovery problem.
To achieve the acquisition of channel parameters, we resort to the Newtonized orthogonal matching pursuit (NOMP) algorithm.
To make the study complete, we also derive the Cramer-Rao lower bound (CRLB).
Finally, the numerical evaluations show the effectiveness of the proposed scheme.

\section{Wideband Channel Model over RIS Assisted Network}

\begin{figure}[!t]
	\centering
	\includegraphics[width=4in]{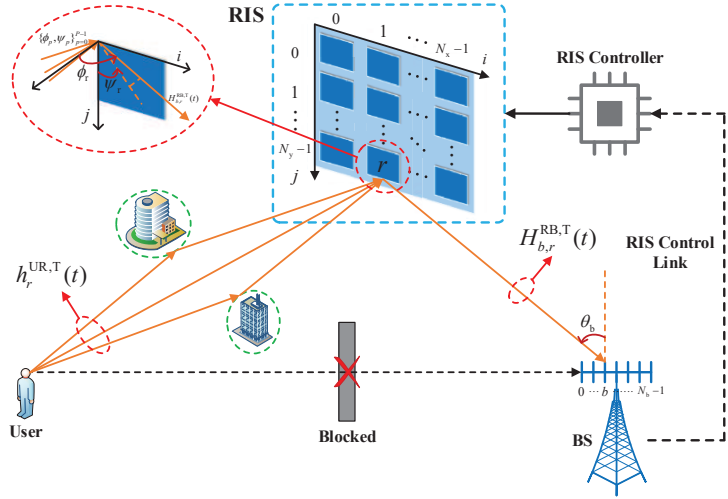}
	\caption{{An RIS assisted mmWave MIMO system.}}
	\label{fig:scene}
\end{figure}

We consider an RIS assisted mmWave MIMO system with a single antenna user, a BS of $N_{\text{b}}$ antennas and an RIS consisting of $N_{\text{r}}$ passive elements, as depicted in Fig. \ref{fig:scene}. The antennas at BS form a uniform linear array (ULA), and the elements in the RIS form a uniform planar array (UPA) with $N_{\text{x}}$ and $N_{\text{y}}$ elements along the horizontal and vertical directions, respectively.
Orthogonal frequency division multiplexing (OFDM) with $N_{\text{c}}$ subcarriers is adopted for combating the multipath delay spread.
Let us denote the transmission bandwidth as $W$, and the subcarrier spacing is $\Delta \!f \!=\!W\!/\!N_{\text{c}}$.
The length of the cyclic prefix is assumed to be longer than that of the maximum multipath delay plus the maximum delay of the antenna domain.

{Suppose that} the multipath channel from the user to RIS consists of $P$ incident paths.
Denote $\tau_{p,r}^{\text{UR}}$ as the time delay of the $p$-th path from the user to the $r$-th element of the RIS, where $p\!\in\!\{0\!,\cdots\!,P\!-\!1\}$ and $r\!\in\!\{0\!,\cdots\!,N_{\text{r}}\!-\!1\}$.
Note that $r=iN_{\text{y}}+j$ where $i\!\in\!\{0\!,\cdots\!,N_{\text{x}}\!-\!1\}$ and $j\!\in\!\{0,\cdots,N_{\text{y}}-1\}$ are the {element indexes along the} horizontal and vertical directions, respectively. {Moreover,} it can be inferred that $j \!= \!(r)_{N_{\text{y}}}$ and $i \!=\! (r\!-\!(r)_{N_{\text{y}}})/N_{\text{y}}$,
where $j\! = \!(r)_{N_{\text{y}}}$ denotes the remainder of $r$ divided by $N_{\text{y}}$.
Denote $\phi_{p}$ and $\psi_{p}$ as the elevation and azimuth {AoAs along}  the $p$-th path at the RIS, respectively.
{With the RIS structure and the far-field assumption \cite{boleiwang}, $\tau_{p,r}^{\text{UR}}$ can be denoted as}
\begin{align}
\tau_{p,r}^{\text{UR}}\!=\!\tau_p\!+\!\frac{d\big(\frac{r-(r)\!_{N_{\text{y}}}}{N_{\text{y}}}\sin(\phi_p)\!\sin(\psi_p)\!\!+\!\!(r)\!_{N_{\text{y}}}\sin(\phi_p)\!\cos(\psi_p)\big)}{c},
\end{align}
{where $d$ is the antenna spacing and $c$ is the speed of light.
Denote the complex path gain of the $p$-th path as $g_{p}^{\text{UR}}$.
Then, the impulse response of the multipath channel from the user to the $r$-th element at the RIS can be expressed as}
\begin{align}
h^{\text{UR},\text{T}}_{r}(t)=\sum_{p=0}^{P-1} \underbrace{g^{\text{UR}}_{p}e^{-\jmath 2 \pi f_{\text{c}} \tau_{p}}}_{\bar{g}^{\text{UR}}_{p}}e^{-\jmath 2 \pi \varpi_{r}(\phi_p,\psi_p)} \delta(t-\tau_{p,r}^{\text{UR}}),
\label{eq:h_RU}
\end{align}
{where $f_{\text{c}}$ is the carrier frequency, $\delta(\cdot)$ is the Delta function, $\bar{g}^{\text{UR}}_{p}$ is the equivalent complex gain, $\varpi_{r}(\phi,\psi)\triangleq\big(d(\frac{r-(r)_{N_{\text{y}}}}{N_{\text{y}}}\sin(\phi)\!\sin(\psi)\!+\!(r)_{N_{\text{y}}}\sin(\phi)\!\cos(\psi))\big)/\lambda_{\text{c}}$ is the normalized angle at the RIS and $\lambda_{\text{c}}$ is the carrier wavelength \cite{boleiwang}.}
%\vspace{-1mm}
%\begin{small}
%\begin{align}
%\varpi_{r}(\phi,\psi)\!\triangleq\! \frac{d(\frac{r-(r)_{N_{\text{y}}}}{N_{\text{y}}}\sin(\phi)\!\sin(\psi)\!+\!(r)_{N_{\text{y}}}\sin(\phi)\!\cos(\psi))}{\lambda_{\text{c}}},
%\label{eq:N_AOA_R}
%\end{align}
%\end{small}where $\lambda_{\text{c}}$ is the carrier wavelength.

The direct path from the user to BS may be blocked by possible obstacles such as buildings and trees \cite{shunbozhang}.
Hence, we mainly focus on the RIS assisted link.
Generally, BS and RIS are considered to be located in an environment with limited local scattering, which would cause the MIMO {link} between BS and RIS to be a light-of-sight (LoS) \cite{angle_known}.
Thus, we define $g^{\text{RB}}$ as the complex path gain, $\theta_{\text{b}}$ as the AoA at BS, $\phi_{\text{r}}/\psi_{\text{r}}$ as the elevation/azimuth AoDs at RIS and $\tau_{b,r}^{\text{RB}} = bd\sin(\theta_{\text{b}})/c+\varpi_{r}(\phi_{\text{r}},\psi_{\text{r}})/f_{\text{c}}$ as the time delay from the $r$-th element of the RIS to the $b$-th antenna of the BS, where $c$ is the speed of light. Then, the impulse response of the LoS channel from the $r$-th {RIS} element to the $b$-th {BS} antenna  can be expressed as
\begin{align}
H_{b,r}^{\text{RB},\text{T}}(t) = g^{\text{RB}}e^{-\jmath 2 \pi f_{\text{c}} b\frac{d\sin(\theta_{\text{b}})}{c}} e^{-\jmath 2 \pi \varpi_{r}(\phi_{\text{r}},\psi_{\text{r}})} \delta(t-\tau_{b,r}^{\text{RB}}).
\label{eq:h_BR}
\end{align}

\section{NOMP based Wideband Channel Estimation for RIS Assisted MIMO}

\subsection{Wideband Effect on Transmission Model}

By taking the Fourier transform of \eqref{eq:h_RU} and stacking it from RIS's different antennas into a $N_{\text{r}}\times1$ vector, the frequency response between the RIS and user can be obtained as
\begin{align}
\mathbf{h}^{\text{UR},\text{F}}(f)=\sum_{p=0}^{P-1} \bar{g}^{\text{UR}}_{p} \mathbf{a}_{\text{R}}(f,\phi_p,\psi_p)e^{-\jmath 2\pi f\tau_{p}},
\end{align}
where $\mathbf{a}_{\text{R}}(f,\phi,\psi)=[1,\cdots,e^{-\jmath 2 \pi (1+\frac{f}{f_{\text{c}}})\varpi_{N_{\text{r}-1}}(\phi,\psi)}]^T$ is the array steering vector at the RIS.
{By a similar mathematical manipulation} on \eqref{eq:h_BR}, the frequency response between the BS and RIS can be expressed as
\begin{align}
\mathbf{H}^{\text{RB},\text{F}}(f) =g^{\text{RB}}\mathbf{a}_{\text{B}}( f,\theta_{\text{b}})\mathbf{a}^T_{\text{R}}(f,\phi_{\text{r}},\psi_{\text{r}}),
\end{align}
where $\mathbf{a}_{\text{B}}(f,\theta)=[1,\cdots,e^{-\jmath 2 \pi (1+\frac{ f}{f_{\text{c}}})(N_{\text{b}}-1)\frac{d\sin(\theta)}{\lambda_{\text{c}}}}]^T$ is the array steering vector at the BS.

Before proceeding, let us denote the RIS phase shift vector at the $m$-th {OFDM} pilot symbol as $\boldsymbol{\rho}_m\triangleq[e^{\jmath\rho_{m,0}},\cdots,e^{\jmath\rho_{m,N_{\text{r}}-1}}]^T$, where $m\in\{0,\cdots,M-1\}$, $M$ is the number of pilot symbols during channel estimation, and $\rho_{m,r}\in[0,2\pi)$ represents the phase shift of the $r$-th RIS element at the $m$-th pilot symbol.
Then, the operation of RIS {can be} described by the diagonal matrix $\boldsymbol{\Omega}_m=\text{diag}\{\boldsymbol{\rho}_m\}\in\mathbb{C}^{N_{\text{r}}\times N_{\text{r}}}$.
Note that $\boldsymbol{\rho}_m$ is assumed to be constant during each pilot symbol.
Then, the frequency response of BS's received signal at the $m$-th pilot symbol can be expressed as
\begin{align}
\mathbf{y}_{m}(f)&=\sum_{p=0}^{P-1} \underbrace{g^{\text{RB}}\bar{g}^{\text{UR}}_{p}}_{g_{p}}\mathbf{a}_{\text{B}}( f,\theta_{\text{b}})\mathbf{a}^T_{\text{R}}(f,\phi_{\text{r}},\psi_{\text{r}}) \boldsymbol{\Omega}_m\mathbf{a}_{\text{R}}(f,\phi_p,\psi_p)\notag\\
&\kern 60pt \times s_{m}(f)e^{-\jmath 2\pi f\tau_{p}}+ \mathbf{v}_{m}(f),
\end{align}
where $g_p$ is defined as the cascaded complex gain, $s_m(f)$ is the frequency-domain signal from the user,
{and} $\mathbf{v}_{m}(f)$ is the additive Gaussian noise with zero mean and covariance matrix $\sigma_{\text v}^2\mathbf I_{N_{\text{b}}}$.
%and  with each element independently distributed as $\mathcal{CN}(0,\sigma_{\text{v}}^2)$.

%\begin{figure*}[!t]
%\vspace{-3mm}
%\begin{small}
%\begin{align}
%&\mathbf{f}(\theta_{\text{b}},\phi_{\text{r}},\psi_{\text{r}},\phi_p,\psi_p,\tau_p)\!=\!\bigg[\Big[\big(\bar{\mathbf{A}}_{l_0}(\phi_{\text{r}},\psi_{\text{r}})\mathbf{a}_{\text{R}}(l_0\Delta f,\phi_p,\psi_p)\big)\!\otimes \!\mathbf{a}_{\text{B}}(l_0\Delta f,\theta_{\text{b}})e^{-\jmath 2\pi l_0\Delta f\tau_{p}}\Big]^T,\cdots,\notag\\
%&\kern 203pt \Big[\big(\bar{\mathbf{A}}_{l_{L-1}}(\phi_{\text{r}},\psi_{\text{r}})\mathbf{a}_{\text{R}}(l_{L-1}\Delta f,\phi_p,\psi_p)\big)\!\otimes \!\mathbf{a}_{\text{B}}(l_{L-1}\Delta f,\theta_{\text{b}})e^{-\jmath 2\pi l_{L-1}\Delta f\tau_{p}}\Big]^T\bigg]^T.
%\label{eq:F}
%\end{align}
%\end{small}
%\vspace{-3mm}
%\begin{small}
%\begin{align}
%\mathbf{f}(\bar{\phi},\bar{\psi},\bar{\tau})\!\!=\!\!\bigg[\Big[\big(\bar{\mathbf{A}}_{l_0}\mathbf{a}_{\text{R}}(l_0\Delta f,\bar{\phi},\bar{\psi})\big)\!\otimes\! \mathbf{a}_{\text{B}}(l_0\Delta f,\theta_{\text{b}})e^{-\jmath 2\pi l_0\Delta f\bar{\tau}}\Big]^T\!\!,\!\cdots\!,\! \Big[\big(\bar{\mathbf{A}}_{l_{L\!-\!1}}\mathbf{a}_{\text{R}}(l_{L\!-\!1}\Delta f,\bar{\phi},\bar{\psi})\big)\!\otimes\! \mathbf{a}_{\text{B}}(l_{L\!-\!1}\Delta f,\theta_{\text{b}})e^{-\jmath 2\pi l_{L\!-\!1}\Delta f\bar{\tau}}\Big]^T\bigg]^T.
%\label{eq:F_grid}
%\end{align}
%\end{small}
%\vspace{-3mm}
%\hrulefill
%\end{figure*}

Assume that $L$ out of $N_{\text{c}}$ subcarriers are exclusively assigned to the user as pilots whose index set is denoted by $\mathcal{L}=\{l_0,\cdots,l_{L-1}\}$.
{Without loss of generality, we assume that the values of all pilots are 1, i.e., $s_m(k\Delta f)=1$, $m\in\{0,\cdots,M-1\}$, $k\in \mathcal L$. }
Then, we collect the received pilot vectors $\mathbf y_0{(k\Delta f)},\cdots, \mathbf y_{M-1}{(k\Delta f)}$ into an $MN_{\text{b}}\times 1$ vector at the $k$-th subcarrier as
%Note that $f = k\Delta f$ and $k$ is the index of subcarrier along the frequency dimension.
%Then, by stacking the received pilots of $M$ blocks into a vector, the received signal vector at the $N_{\text{b}}$ antennas of the BS in the $k$-th subcarrier with $M$ pilot blocks can be expressed as
\begin{align}
\mathbf y(k\Delta f) &\triangleq [\mathbf{y}_{0}^T(k\Delta f),\cdots,\mathbf{y}_{M-1}^T(k\Delta f)]^T\notag\\
&=\sum_{p=0}^{P-1}\! g_p\!\Big(\!\underbrace{\left[\begin{array}{c}
\!\!\!\mathbf{a}^T_{\text{R}}(k\Delta f,\phi_{\text{r}},\psi_{\text{r}}) \boldsymbol{\Omega}_0\!\!\! \\
\vdots \\
\!\!\!\mathbf{a}^T_{\text{R}}(k\Delta f,\phi_{\text{r}},\psi_{\text{r}}) \boldsymbol{\Omega}_{M-1}\!\!\!
\end{array}\right]}_{\bar{\mathbf{A}}_k(\phi_{\text{r}},\psi_{\text{r}})\in\mathbb{C}^{M\times N_{\text{r}}}}\!\mathbf{a}_{\text{R}}(k\Delta f,\phi_p,\psi_p)\!\Big)\notag\\
&\kern 10pt\otimes \mathbf{a}_{\text{B}}(k\Delta f,\theta_{\text{b}})e^{-\jmath 2\pi k\Delta f\tau_{p}}\!+\!\mathbf{v}(k\Delta f),\quad k\in \mathcal L,
\end{align}
where {$\bar{\mathbf{A}}_k(\phi_{\text{r}},\psi_{\text{r}})$ is defined above} and $\mathbf{v}(k\Delta f)=[\mathbf v_0^T(k\Delta f),\cdots,
\mathbf v_{M-1}^T(k\Delta f)]^T$ is {the} noise vector.
Define $\mathbf{f}(\theta_{\text{b}},\phi_{\text{r}},\psi_{\text{r}},\phi_p,\psi_p,\tau_p)\in\mathbb{C}^{LMN_{\text{b}}\times1}$ as
\begin{align}
&\mathbf{f}(\theta_{\text{b}},\phi_{\text{r}},\psi_{\text{r}},\phi_p,\psi_p,\tau_p)\!=\!\bigg[\Big[\big(\bar{\mathbf{A}}_{l_0}(\phi_{\text{r}},\psi_{\text{r}})\mathbf{a}_{\text{R}}(l_0\Delta f,\phi_p,\psi_p)\big)\!\otimes \!\mathbf{a}_{\text{B}}(l_0\Delta f,\theta_{\text{b}})e^{-\jmath 2\pi l_0\Delta f\tau_{p}}\Big]^T,\cdots,\notag\\
&\kern 88pt \Big[\big(\bar{\mathbf{A}}_{l_{L-1}}(\phi_{\text{r}},\psi_{\text{r}})\mathbf{a}_{\text{R}}(l_{L-1}\Delta f,\phi_p,\psi_p)\big)\!\otimes \!\mathbf{a}_{\text{B}}(l_{L-1}\Delta f,\theta_{\text{b}})e^{-\jmath 2\pi l_{L-1}\Delta f\tau_{p}}\Big]^T\bigg]^T.
\label{eq:F}
\end{align}
By collecting $\mathbf y(k\Delta f)$ at different subcarriers, we obtain
\begin{align}
\mathbf{y}&\triangleq[\mathbf{y}^T(l_0\Delta f),\cdots,\mathbf{y}^T(l_{L-1}\Delta f)]^T\notag\\
&=\sum_{p=0}^{P-1} g_p\mathbf{f}(\theta_{\text{b}},\phi_{\text{r}},\psi_{\text{r}},\phi_p,\psi_p,\tau_p)+\mathbf{v},
\label{eq:y_final}
\end{align}
where $\mathbf{v}\triangleq[\mathbf{v}^T(l_0\Delta f),\cdots,\mathbf{v}^T(l_{L-1}\Delta f)]^T\in\mathbb{C}^{LMN_{\text{b}}\times1}$ is the corresponding noise vector.

We are interested in estimating the cascaded channels at the BS from \eqref{eq:y_final}, which can be transformed into the problem of parameter recovery.
Without loss of generality, we assume that the locations of the BS and RIS are known, which implies that {the angles $\theta_{\text{b}}$, $\phi_{\text{r}}$ and $\psi_{\text{r}}$} can be completely determined by the geometric positions of BS and RIS.
Then, $\{\theta_{\text{b}},\phi_{\text{r}},\psi_{\text{r}}\}$ are constants and can be omitted in $\bar{\mathbf{A}}_{k}(\phi_{\text{r}},\psi_{\text{r}})$ and  $\mathbf{f}(\theta_{\text{b}},\phi_{\text{r}},\psi_{\text{r}},\phi_p,\psi_p,\tau_p)$.
In order to estimate the RIS assisted channel, we will resort to the NOMP algorithm to capture the parameter set $\{g_p,\phi_{\text{p}},\psi_{\text{p}},\tau_{p}\}_{p=0}^{P-1}$.

\subsection{NOMP Algorithm}
The NOMP algorithm is divided into five steps: Greedy searching, precise searching, single refinement, cyclic refinement and cascaded gain updating.
\begin{enumerate}
\item {\it{Greedy Searching}}:
The over-sampled grids along the horizontal AoA, vertical AoA and delay are respectively divided with sampling rate $\eta_{\phi}$, $\eta_{\psi}$ and $\eta_{\tau}$, where the range of actual horizontal AoA, vertical AoA and time delay are respectively $[-\pi/2,\pi/2)$, $[0,\pi/2)$ and $[0,1/\Delta f)$.
Hence, the codeword corresponding to \eqref{eq:F} is expressed as
\begin{align}
\mathbf{f}(\bar{\phi},\bar{\psi},\bar{\tau})\!&=\!\bigg[\Big[\big(\bar{\mathbf{A}}_{l_0}\mathbf{a}_{\text{R}}(l_0\Delta f,\bar{\phi},\bar{\psi})\big)\!\otimes\! \mathbf{a}_{\text{B}}(l_0\Delta f,\theta_{\text{b}})e^{-\jmath 2\pi l_0\Delta f\bar{\tau}}\Big]^T\!\!,\!\cdots\!,\!\notag\\
&\kern 50pt \Big[\big(\bar{\mathbf{A}}_{l_{L\!-\!1}}\mathbf{a}_{\text{R}}(l_{L\!-\!1}\Delta f,\bar{\phi},\bar{\psi})\big)\!\otimes\! \mathbf{a}_{\text{B}}(l_{L\!-\!1}\Delta f,\theta_{\text{b}})e^{-\jmath 2\pi l_{L\!-\!1}\Delta f\bar{\tau}}\Big]^T\bigg]^T.
\label{eq:F_grid}
\end{align}
where $\bar{\phi}\in\{-\frac{\pi}{2},\cdots,-\frac{\pi}{2}+\frac{(\eta_{\phi}N_{\text{x}}-1)\pi}{\eta_{\phi}N_{\text{x}}}\}$, $\bar{\psi}\in\{0,\cdots,\frac{(\eta_{\psi}N_{\text{y}}-1)\pi}{2\eta_{\psi}N_{\text{y}}}\}$ and $\bar{\tau}\in\{0,\cdots,\frac{\eta_{\tau}N_{\text{c}}-1}{\eta_{\tau}N_{\text{c}}\Delta f}\}$.

At the beginning of the $i$-th iteration, the residual noisy mixture $\mathbf{y}_{\text{e}}^i$ is calculated by
\begin{align}
\mathbf{y}_{\text{e}}^{i}=\mathbf{y}-\sum_{p=0}^{i-1} \hat{g}_{p} \mathbf{f}(\hat{\phi}_p,\hat{\psi}_p,\hat{\tau}_p),
\end{align}
where $\{\hat{g}_{p},\hat{\phi}_p,\hat{\psi}_p,\hat{\tau}_p\}_{p=0}^{i-1}$ are the estimated parameters in the previous iterations.
The course estimated $\hat{\phi}_i$, $\hat{\psi}_i$, $\hat{\tau}_i$ are obtained by greedily searching the grid points as
\begin{align}
(\hat{\phi}_i,\hat{\psi}_i,\hat{\tau}_i)=\underset{(\bar{\phi},\bar{\psi},\bar{\tau})}{\arg\max}\frac{|\mathbf{f}^H(\bar{\phi},\bar{\psi},\bar{\tau})\mathbf{y}_{\text{e}}^{i}|^2}{\|\mathbf{f}(\bar{\phi},\bar{\psi},\bar{\tau})\|^2},
\label{eq:f_coarse}
\end{align}
where $\hat{\phi}_i$, $\hat{\psi}_i$ and $\hat{\tau}_i$ are the estimated horizontal AoA, vertical AoA and time delay of the $i$-th component path, respectively.

\item {\it{Precise Searching}}:
After the simultaneously greedy searching for the three parameters, we perform a precise searching process near the results of the greedy searching step. Notice that the implementation of this process is similar to that of the greedy searching step and the details are omitted here {due to space limitation}. Then, the estimated parameters $\{\hat{\phi}_i,\hat{\psi}_i,\hat{\tau}_i\}$ can be obtained through a similar operation in \eqref{eq:f_coarse}.
Afterwards, the coarse estimation of the cascaded gain $\hat{g}_{i}$ can be obtained as
\begin{align}
\hat{g}_{i}=\frac{\mathbf{f}^H(\hat{\phi}_i,\hat{\psi}_i,\hat{\tau}_i)\mathbf{y}_{\text{e}}^{i}}{\|\mathbf{f}(\hat{\phi}_i,\hat{\psi}_i,\hat{\tau}_i)\|^2}.
\label{eq:g_coarse}
\end{align}

\item {\it{Single Refinement}}:
We will resort to the extended Newton method and refine $\hat{\phi}_i$, $\hat{\psi}_i$, $\hat{\tau}_i$ and $\hat{g}_{i}$, and $R_{\text{s}}$ iterations are executed in this step.
The goal of the refinement step is to minimize the power of the new residual $\|\mathbf{y}_{\text{e}}^i-g\mathbf{f}(\phi,\psi,\tau)\|^2$ in the $i$-th iteration.
Hence, the target is to maximize
\begin{align}
\mathcal{N}(\phi,\psi,\tau) \!=\! 2\mathcal{R}\{(\mathbf{y}_{\text{e}}^i)^Hg\mathbf{f}(\phi,\psi,\tau)\}\!-\!\|g\mathbf{f}(\phi,\psi,\tau)\|^2.
\label{eq:target_N}
\end{align}
Then, the refined estimations of $\hat{\phi}_i$, $\hat{\psi}_i$ and $\hat{\tau}_i$ can be expressed as
\begin{align}
\left[\begin{array}{l}
\!\!\hat{\phi}_{i}\!\!\!\! \\
\!\!\hat{\psi}_{i}\!\!\!\! \\
\!\!\hat{\tau}_{i}\!\!\!\!
\end{array}\right]\!\!=\!\!\left[\begin{array}{l}
\!\!\hat{\phi}_{i}\!\!\!\! \\
\!\!\hat{\psi}_{i}\!\!\!\! \\
\!\!\hat{\tau}_{i}\!\!\!\!
\end{array}\right]\!\!\!-\!\!\ddot{\boldsymbol{\mathcal{N}}}^{-1}\!\left(\hat{g}_{i}, \hat{\phi}_{i}, \hat{\psi}_{i},\hat{\tau}_{i}\right) \dot{\boldsymbol{\mathcal{N}}}\!\left(\hat{g}_{i}, \hat{\phi}_{i}, \hat{\psi}_{i},\hat{\tau}_{i}\right),
\label{eq:Newton}
\end{align}
where
$\dot{\boldsymbol{\mathcal{N}}}\left(g, \phi, \psi, \tau \right)=[
\frac{\partial \mathcal{N}}{\partial \phi},\frac{\partial \mathcal{N}}{\partial \psi},
\frac{\partial \mathcal{N}}{\partial \tau}]^T$
is the first-order partial derivative vector, and
\begin{align}
\ddot{\boldsymbol{\mathcal{N}}}(g, \phi, \psi, \tau)=\left[\begin{array}{lll}
\frac{\partial^{2} \mathcal{N}}{\partial \phi^{2}} & \frac{\partial^{2} \mathcal{N}}{\partial \phi \partial \psi} & \frac{\partial^{2} \mathcal{N}}{\partial \phi \partial \tau}\\
\frac{\partial^{2} \mathcal{N}}{\partial \psi \partial \phi} & \frac{\partial^{2} \mathcal{N}}{\partial \psi^{2}} & \frac{\partial^{2} \mathcal{N}}{\partial \psi \partial \tau}\\
\frac{\partial^{2} \mathcal{N}}{\partial \tau \partial \phi} & \frac{\partial^{2} \mathcal{N}}{\partial \tau \partial \psi}& \frac{\partial^{2} \mathcal{N}}{\partial \tau^2}
\end{array}\right]
\end{align}
is the second-order partial derivative matrix. According to \eqref{eq:target_N}, we can write the first-order partial derivatives of $\mathcal{N}(\phi,\psi,\tau)$ as
$\frac{\partial \mathcal{N}}{\partial x}=2 \mathcal{R}\left\{g\left(\mathbf{y}_{\text{e}}^i-g \mathbf{f}\right)^{H} \frac{\partial \mathbf{f}}{\partial x}\right\}$,
where $x$ can be $\phi$, $\psi$ and $\tau$. The second-order partial derivative of $\mathcal{N}(\phi,\psi,\tau)$ can be calculated as
$\frac{\partial^{2} \mathcal{N}}{\partial x_{1} \partial x_{2}}=2 \mathcal{R}\left\{g\left(\mathbf{y}_{\text{e}}^i-g \mathbf{f}\right)^{H} \frac{\partial^{2} \mathbf{f}}{\partial x_{1} \partial x_{2}}-\left|g\right|^{2} \frac{\partial \mathbf{f}^{H}}{\partial x_{2}} \frac{\partial \mathbf{f}}{ \partial x_{1}}\right\}$,
where $x_1$ and $x_2$ can be $\phi$, $\psi$ and $\tau$.
Due to the space limitation, the partial derivatives are omitted here.

By carrying out \eqref{eq:Newton}, $\{\hat{\phi}_i,\hat{\psi}_i,\hat{\tau}_i\}$ are refined.
Then, the cascaded gain is also updated according to \eqref{eq:g_coarse}.

\item {\it{Cyclic Refinement}}:
After the single refinement step for the parameters of the current iteration, $R_{\text{c}}$ iterations of cyclically refinement are taken into consideration to further perfect the estimations $\{\hat{g}_i,\hat{\phi}_i,\hat{\psi}_i,\hat{\tau}_i\}_{p=0}^{i-1}$ of the previous iterations.
Similar to the single refinement step, the extended Newton method is also utilized, and the accurate estimations $\{\hat{g}_i$, $\hat{\phi}_i$, $\hat{\psi}_i, \hat{\tau}_i\}_{p=0}^{i}$ can be obtained.
The derivation is omitted due to space limitation.

\item {\it{Cascaded Gain Updating}}:
Based on $\{\hat{\phi}_p,\hat{\psi}_p,\hat{\tau}_p\}_{p=0}^{i}$ obtained in the previous iterations, we can further update the cascaded gains {through the least square algorithm} as
\begin{align}
[\hat{g}_{0},\cdots,\hat{g}_{i}]^T = (\mathbf{F}_i^H\mathbf{F}_i)^{-1}\mathbf{F}_i^H\mathbf{y},
\end{align}
where $\mathbf{F}_i=[\mathbf{f}(\hat{\phi}_0,\hat{\psi}_0,\hat{\tau}_0),\cdots,\mathbf{f}(\hat{\phi}_i,\hat{\psi}_i,\hat{\tau}_i)]$.
\end{enumerate}

%In fact, since the BS does not know the number of propagation paths in the real channel, a stopping criterion {should} be designed to determine when the iteration process is terminated.
As the NOMP runs, the power of the residual $\mathbf{y}_{\text{e}}^i$ decreases after each iteration. If the extracted parameters are accurate enough, the power of the residual is reduced to the power of the noise in the end, i.e., $\|\mathbf{y}^i_{\text{e}}\|^2\approx\|\mathbf{v}\|^2$.
Hence, the NOMP algorithm terminates when
$|\mathbf{f}^H(\bar{\phi}, \bar{\psi}, \bar{\tau})\mathbf{y}_{\text{e}}^i|^2<\epsilon$
for all possible $(\bar{\phi}, \bar{\psi}, \bar{\tau})$.
The stopping criterion threshold $\epsilon$ is chosen from the false alarm rate $P_{\text{fa}}\!=\!P\{\|\mathbf{v}\|^2_{\infty}\!>\!\epsilon\}\!=\!1\!-\!(1\!-\!\exp(-\!\epsilon/\sigma^2_{\text{v}}))^{LMN_{\text{b}}}$ as $\epsilon\!=\!-\sigma_{\text{v}}^{2} \ln (1\!-\!(1\!-\!P_{\text{fa}})^{1 / LMN_{\text{b}}})$ \cite{NOMP}.

\vspace{-3mm}
\subsection{Performance Analysis}
%After using the NOMP algorithm, we assume that there are the parameters $\{\hat{g}_p,\hat{\phi}_p, \hat{\psi}_p, \hat{\tau}_p\}_{p=0}^{P-1}$ of $P$ paths are extracted, and the corresponding channel can be constructed.
{To evaluate the performance of the NOMP algorithm, we derive the CRLB.}
%Then, we will analyze the lower bound of the estimation regarding these four parameters.
{Let us define the $5P\times 1$ unknown parameter vector} $\boldsymbol{\xi}=[\boldsymbol{\xi}^T_1,\cdots,\boldsymbol{\xi}^T_{P}]^T$, where
$\boldsymbol{\xi}_p=[|g_p|,\angle g_p,\phi_p,\psi_p,\tau_p]^T\in\mathbb{R}^{5\times1}$
{denotes the unknown} parameters of the $p$-th path, and $|g_p|$, $\angle g_p$ {separately represent} the amplitude and phase of the complex-valued path gain.
Then, for an unbiased estimator, {the} estimation variance is bounded by CRLB, which is the inverse of the $5P\times 5P$ Fisher information matrix (FIM) $\boldsymbol{\mathcal{F}}(\boldsymbol{\xi})$ \cite{FIM} defined as
\begin{align}
[\boldsymbol{\mathcal{F}}(\boldsymbol{\xi})]_{i, j}=\mathbb{E}_{\mathbf{y} \mid \boldsymbol{\xi}}\left\{\frac{\partial \ln p(\mathbf{y} \mid \boldsymbol{\xi})}{\partial \xi_{i}} \frac{\partial \ln p(\mathbf{y} \mid \boldsymbol{\xi})}{\partial \xi_{j}}\right\},
\end{align}
where $p(\mathbf{y} \mid \boldsymbol{\xi})$ is the likelihood function of $\mathbf{y}$ conditioned on $\boldsymbol{\xi}$, and {the} expectation is taken over {the} noise distribution.
The $5P\times 5P$ FIM can be sliced into $P^2$ submatrices as
\begin{align}
\boldsymbol{\mathcal{F}}(\boldsymbol{\xi})=\left[\begin{array}{ccc}
\boldsymbol{\mathcal{F}}(\boldsymbol{\xi}_{1}, \boldsymbol{\xi}_{1}) & \cdots & \boldsymbol{\mathcal{F}}(\boldsymbol{\xi}_{1}, \boldsymbol{\xi}_{P}) \\
\vdots & \ddots & \vdots \\
\boldsymbol{\mathcal{F}}(\boldsymbol{\xi}_{P}, \boldsymbol{\xi}_{1}) & \cdots & \boldsymbol{\mathcal{F}}(\boldsymbol{\xi}_{P}, \boldsymbol{\xi}_{P})
\end{array}\right],
\end{align}
where $\boldsymbol{\mathcal{F}}(\boldsymbol{\xi}_{i}, \boldsymbol{\xi}_{j})$ is a $5\times 5$ matrix.
Generally, a closed-form CRLB analysis for multipath channel estimation is hard to obtain \cite{CRLB_gao}.
Moreover, if two or more paths have an extremely close angle and delay, the rank deficiency of $\boldsymbol{\mathcal{F}}(\boldsymbol{\xi})$ appears, which causes the determinant of $\boldsymbol{\mathcal{F}}(\boldsymbol{\xi})$ to be close to $0$.
However, the number of paths to the BS is very small and these paths are characterized by separation in delay in wideband mmWave systems.
Hence,
we assume that $\boldsymbol{\mathcal{F}}(\boldsymbol{\xi})$ is nonsingular and can be transformed into a block diagonal matrix \cite{FIM}, and each submatrix on the diagonal of $\boldsymbol{\mathcal{F}}(\boldsymbol{\xi})$ can be written as
\begin{align}
&\boldsymbol{\mathcal{F}}(\!\boldsymbol{\xi}\!_{p}\!, \! \boldsymbol{\xi}\!_{p}\!)
=\!\!\left[\begin{array}{lllll}
\!\!\!\boldsymbol{\mathcal{F}}(\!|g\!_{p}\!|\!,\!|g\!_{p}\!|\!) \!\!\!& \!\! \! \boldsymbol{\mathcal{F}}(\!|g\!_{p}\!|, \!\angle\! g\!_{p}\!) \!\!\!&\!\! \! \boldsymbol{\mathcal{F}}(\!|g_{p}\!|,\! \phi\!_{p}\!)\! \!\!& \!\! \! \boldsymbol{\mathcal{F}}(\!|g_{p}\!|, \!\psi\!_{p}\!) \!\!\!& \!\! \! \boldsymbol{\mathcal{F}}(\!|g_{p}\!|,\! \tau\!_{p}\!)\!\!\! \\
\!\!\!\boldsymbol{\mathcal{F}}(\!\angle\! g\!_{p}\!,\!|g\!_{p}\!|\!) \!\!\!&\! \!\! \boldsymbol{\mathcal{F}}(\!\angle \!g\!_{p}\!,\! \angle\! g\!_{p}\!) \!\!\!&\! \!\! \boldsymbol{\mathcal{F}}(\!\angle\! g\!_{p}\!,\! \phi\!_{p}\!) \!\!\!&\! \!\! \boldsymbol{\mathcal{F}}(\!\angle\! g\!_{p}\!,\! \psi\!_{p}\!) \!\!\!&\! \!\! \boldsymbol{\mathcal{F}}(\!\angle\! g\!_{p}\!,\! \tau\!_{p}\!) \!\!\!\\
\!\!\!\boldsymbol{\mathcal{F}}(\!\phi\!_{p}\!,\!|g\!_{p}\!|\!) \!\!\!&\! \!\! \boldsymbol{\mathcal{F}}(\!\phi\!_{p}\!,\! \angle\! g\!_{p}\!) \!\!\!&\! \!\! \boldsymbol{\mathcal{F}}(\!\phi\!_{p}\!,\! \phi\!_{p}\!)\!\!\!&\! \!\! \boldsymbol{\mathcal{F}}(\!\phi\!_{p}\!,\! \psi\!_{p}\!)\!\!\!&\! \!\! \boldsymbol{\mathcal{F}}(\!\phi\!_{p}\!, \!\tau\!_{p}\!)\!\!\! \\
\!\!\!\boldsymbol{\mathcal{F}}(\!\psi\!_{p}\!,\!|g\!_{p}\!|\!) \!\!\!&\! \!\! \boldsymbol{\mathcal{F}}(\!\psi\!_{p}\!,\! \angle\! g\!_{p}\!) \!\!\!& \!\!\! \boldsymbol{\mathcal{F}}(\!\psi\!_{p}\!,\! \phi\!_{p}\!) \!\!\!& \!\!\! \boldsymbol{\mathcal{F}}(\!\psi\!_{p}\!,\! \psi\!_{p}\!) \!\!\!&\! \!\! \boldsymbol{\mathcal{F}}(\!\psi\!_{p}\!,\! \tau\!_{p}\!)\!\!\! \\
\!\!\!\boldsymbol{\mathcal{F}}(\!\tau_{p}\!,\!|g_{p}\!|\!) \!\!\!&\! \!\! \boldsymbol{\mathcal{F}}(\!\tau\!_{p}\!,\! \angle\! g\!_{p}\!) \!\!\!&\! \!\! \boldsymbol{\mathcal{F}}(\!\tau\!_{p}\!,\! \phi\!_{p}\!) \!\!\!&\! \!\! \boldsymbol{\mathcal{F}}(\!\tau\!_{p}\!, \!\psi\!_{p}\!) \!\!\!& \!\!\! \boldsymbol{\mathcal{F}}(\!\tau\!_{p}\!, \!\tau\!_{p}\!)\!\!\!
\end{array}\right].
\label{eq:F_pa_pa}
\end{align}
Considering that the inverse of a block diagonal matrix is a block diagonal matrix with the inverse of the original blocks on its diagonal, we can separately calculate the inverse matrix of each submatrix. Thus, $\boldsymbol{\mathcal{F}}(\boldsymbol{\xi}_{p}, \boldsymbol{\xi}_{p})$ can be derived as in Appendix and the FIM $\boldsymbol{\mathcal{F}}(\boldsymbol{\xi})$ can be obtained.

%the MSE of the estimated CSI $\widehat{\widetilde{H}}_{k,m,b}$ of the actual cascaded channel $\widetilde{H}_{k,m,b}$ is defined as $\text{MSE}_{k,m,b}(\boldsymbol{\xi})=\mathbb{E}\{|\widehat{\widetilde{H}}_{k,m,b}-\widetilde{H}_{k,m,b}|^{2}\}$,

Let us define the channel statement information (CSI) as
\begin{align}
\widetilde{H}_{k,m,b} = \sum_{p=0}^{P-1}|g_p|e^{\jmath\angle g_p}\breve{A}_{k,m,b,p}e^{-\jmath 2\pi k\Delta f\tau_{p}}.
\end{align}
From \cite{CRLB_gao},
the lower bound of the CSI can be obtained via the transformation vector $\partial \widetilde{H}_{k,m,b}/\partial \boldsymbol{\xi}$ as
\begin{align}
\text{LB}_{k,m,b}(\boldsymbol{\xi})=\left(\frac{\partial \widetilde{H}_{k,m,b}}{\partial \boldsymbol{\xi}}\right)^{H} \boldsymbol{\mathcal{F}}^{-1}(\boldsymbol{\xi}) \frac{\partial \widetilde{H}_{k,m,b}}{\partial \boldsymbol{\xi}},
\label{eq:CSI_CRLB}
\end{align}
where the entries of $\partial \widetilde{H}_{k,m,b}/\partial \boldsymbol{\xi}$ are given as
\begin{align}
\frac{d \widetilde{H}_{k,m,b}}{d |g_p|}\!\!&= \!\!e^{\jmath\angle g_p}\breve{A}_{k,m,b,p}e^{-\jmath 2\pi k\Delta f\tau_{p}},\quad
\frac{d \widetilde{H}_{k,m,b}}{d \angle g_p}\!\!=\!\!\jmath|g_p|e^{\jmath\angle g_p}\breve{A}_{k,m,b,p}e^{-\jmath 2\pi k\Delta f\tau_{p}},\notag\\
\frac{d \widetilde{H}_{k,m,b}}{d \phi_p}\!\!&=\!\!|g_p|e^{\jmath\angle g_p}e^{-\jmath 2\pi k\Delta f\tau_{p}}\frac{\partial \breve{A}_{k,m,b,p}}{\partial \phi_p},\quad
\frac{d \widetilde{H}_{k,m,b}}{d \psi_p}\!\!=\!\!|g_p|e^{\jmath\angle g_p}e^{-\jmath 2\pi k\Delta f\tau_{p}}\frac{\partial \breve{A}_{k,m,b,p}}{\partial \psi_p},\notag\\
\frac{d \widetilde{H}_{k,m,b}}{d \tau_p}\!\!&=\!\!-\jmath2\pi k\Delta f|g_p|e^{\jmath\angle g_p}\breve{A}_{k,m,b,p}e^{-\jmath 2\pi k\Delta f\tau_{p}}.
\label{eq:partH}
\end{align}
Finally, the CSI error \eqref{eq:CSI_CRLB} can be constructed by solving the inverse of $\boldsymbol{\mathcal{F}}(\boldsymbol{\xi})$, where the inversion process of $\boldsymbol{\mathcal{F}}(\boldsymbol{\xi})$ is omitted due to space limitations.
%\begin{figure*}[!t]
%\vspace{-3mm}
%\begin{small}
%\begin{align}
%\frac{d \widetilde{H}_{k,m,b}}{d |g_p|}\!\!&= \!\!e^{\jmath\angle g_p}\breve{A}_{k,m,b,p}e^{-\jmath 2\pi k\Delta f\tau_{p}},\quad
%\frac{d \widetilde{H}_{k,m,b}}{d \angle g_p}\!\!=\!\!\jmath|g_p|e^{\jmath\angle g_p}\breve{A}_{k,m,b,p}e^{-\jmath 2\pi k\Delta f\tau_{p}},\quad
%\frac{d \widetilde{H}_{k,m,b}}{d \phi_p}\!\!=\!\!|g_p|e^{\jmath\angle g_p}e^{-\jmath 2\pi k\Delta f\tau_{p}}\frac{\partial \breve{A}_{k,m,b,p}}{\partial \phi_p},\notag\\
%\frac{d \widetilde{H}_{k,m,b}}{d \psi_p}\!\!&=\!\!|g_p|e^{\jmath\angle g_p}e^{-\jmath 2\pi k\Delta f\tau_{p}}\frac{\partial \breve{A}_{k,m,b,p}}{\partial \psi_p},\quad
%\frac{d \widetilde{H}_{k,m,b}}{d \tau_p}\!\!=\!\!-\jmath2\pi k\Delta f|g_p|e^{\jmath\angle g_p}\breve{A}_{k,m,b,p}e^{-\jmath 2\pi k\Delta f\tau_{p}}.
%\label{eq:partH}
%\end{align}
%\end{small}
%\vspace{-3mm}
%\hrulefill
%\end{figure*}

\section{Simulation Results}

%\begin{figure*}
% \begin{minipage}[t]{60mm}
% \centering
% \includegraphics[width=62mm]{diT.eps}
% \caption{NMSEs of cascaded gain, angle and time delay versus SNR at different $T$.}
% \label{fig:diT}
% \end{minipage}
% \begin{minipage}[t]{60mm}
% \centering
% \includegraphics[width=58mm]{diNL.eps}
% \caption{NMSEs of cascaded gain, angle and time delay versus SNR at different $L$ and $N_{\text{r}}$.}
% \label{fig:diNbNr}
% \end{minipage}
% \begin{minipage}[t]{60mm}
% \centering
% \includegraphics[width=64mm]{channel_H_nmse.eps}
% \caption{Comparison of cascaded channels' NMSEs of OMP, NOMP and CRLB versus SNR.}
% \label{fig:channel_H_nmse}
% \end{minipage}
%\end{figure*}

In this section, we evaluate the performance of the proposed scheme for the wideband channel estimation over RIS assisted mmWave MIMO.
{The system parameter settings are as follows.} The carrier frequency is $f_{\text{c}}=28$ GHz, the bandwidth is $W=600$ MHz, and the number of subcarriers is $N_{\text{c}}=512$.
The RIS assisted mmWave channel parameter settings are: the number of antennas at BS is $N_{\text{b}}=64$, the antenna spacing is $d=\lambda_{\text{c}}/2$, the number of paths is $P=5$, and $\tau_p$, $\phi_p$ and $\psi_p$ are uniformly distributed within $[0,32/W)$, $[-\pi/2,\pi/2)$ and $[-\pi/2,\pi/2)$, respectively.
Moreover, during the channel estimation, the under-sampling rates in the greedy searching are set as $\eta_{\phi}=2$, $\eta_{\psi}=2$ and $\eta_{\tau}=2$, the over-sampling rates in the precise searching are set as $\eta_{\phi}=4$, $\eta_{\psi}=4$ and $\eta_{\tau}=4$, and the iteration number of the single refinement and cyclic refinement {are set as $R_\text{s}=R_\text{c}=5$.}

\begin{figure}[!t]
	\centering
	\includegraphics[width=4in]{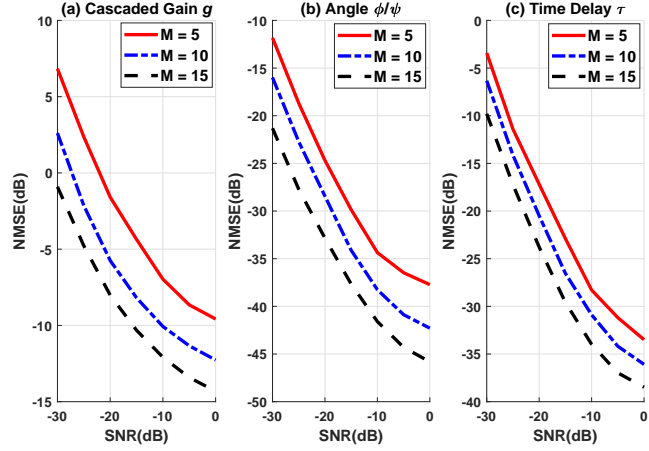}
	\caption{{NMSEs of channel parameters versus SNR at different $M$.}}
	\label{fig:diT}
\end{figure}

\begin{figure}[!t]
	\centering
	\includegraphics[width=4in]{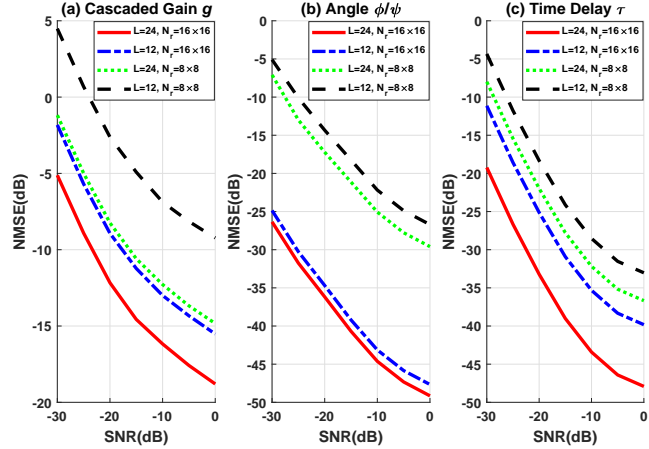}
	\caption{{NMSEs of channel parameters versus SNR at different $L$ and $N_{\text{r}}$.}}
	\label{fig:diNbNr}
\end{figure}

The signal-to-noise ratio (SNR) is expressed as $10\log_{10}\sigma_{\text{t}}^2/\sigma_{\text{v}}^2$, where $\sigma_{\text{t}}^2$ is the average power of the effective signal.
Here, we use the normalized mean square error (NMSE) for the channel parameters as performance metric, which is defined as
$\text{NMSE}_{\mathbf{x}}\!\!\!=\!\!\!\mathbb{E}\left\{\frac{\|\hat{\mathbf{x}}-\mathbf{x}\|^{2}}{\|\mathbf{x}\|^{2}}\right\}$
with $\hat{x}$ being the estimate of $x$ and the $p$-th element of $x$ consisting of $\{g_{p}, \phi_{p}, \psi_{p}, \tau_{p}\}$.
Note that $\mathbf{g}\!\!=\!\![g_0,\!\cdots\!,g_{P-1}]^T$, $\boldsymbol{\phi}\!\!=\!\![\phi_0,\!\cdots\!,\phi_{P-1}]^T$, $\boldsymbol{\psi}\!\!=\!\![\psi_0,\!\cdots\!,\psi_{P-1}]^T$, $\boldsymbol{\tau}\!\!=\!\![\tau_0,\!\cdots\!,\tau_{P-1}]^T$, and {the NMSE of `Angle $\phi/\psi$' in Fig. \ref{fig:diT} and Fig. \ref{fig:diNbNr} denotes the average of the angle $\boldsymbol{\phi}$'s NMSE and $\boldsymbol{\psi}$'s NMSE.}

Fig. \ref{fig:diT} plots the NMSEs of channel parameters estimates with respect to SNR, where {the number of pilot symbols is $M=5$, $10$, $15$}, respectively.
Notice that the number of pilot carriers and elements at RIS are fixed to $L=12$ and $N_{\text{r}}=16\times16$, respectively.
It can be seen from Fig. \ref{fig:diT} that all NMSEs decrease as the SNR increases, and the proposed scheme achieves a good estimation performance even at low SNR, which demonstrates its effectiveness.
Moreover, for the same SNR, the NMSE decreases with {the increase of $M$}, which shows that more pilot symbols can enhance the parameter recovery performance.

Fig. \ref{fig:diNbNr} plots the NMSEs of channel parameters estimates {versus SNR for different $L$ and $N_{\text{r}}$, {where $M=16$}.
It can be seen from Fig. \ref{fig:diNbNr} that all NMSEs of the parameters decrease as the SNR increases.
At the same SNR, the performance of parameter estimation becomes better as the number of RIS elements or pilot subcarriers increases as expected.}

\begin{figure}[!t]
	\centering
	\includegraphics[width=4in]{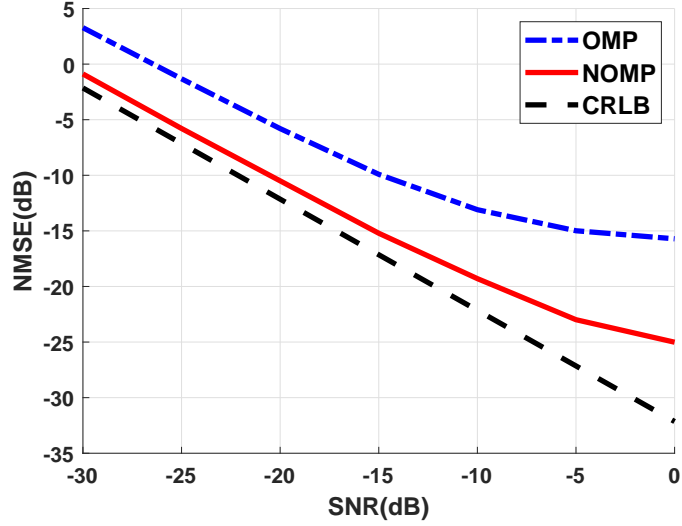}
	\caption{Comparison of cascaded channels' NMSEs of OMP, NOMP and CRLB versus SNR.}
	\label{fig:channel_H_nmse}
\end{figure}

In Fig. \ref{fig:channel_H_nmse}, we employ the NMSE of estimated cascaded channels, i.e.,
$\text{NMSE}_{\widetilde{\mathbf{H}}}\!=\!\mathbb{E}\Big\{\!\frac{\|\widehat{\widetilde{\mathbf{H}}}\!-\!\widetilde{\mathbf{H}}\|^{2}}{\|\widetilde{\mathbf{H}}\|^{2}}\!\Big\}$,
as performance indicator with $\widehat{\widetilde{H}}_{k,m,b}$ {as the estimate} of $\widetilde{H}_{k,m,b}$, where $\widetilde{\mathbf{H}}\!=\![\widetilde{H}_{0,0,0},\cdots,\widetilde{H}_{L\!-\!1,M\!-\!1,N_{\text{b}}\!-\!1}]^T$.
Fig. \ref{fig:channel_H_nmse} compares {the cascaded channel NMSEs of} the OMP method \cite{OMP}, the NOMP method and CRLB at different SNRs, where {$M\!=\!16$}, $L\!=\!12$ and $N_{\text{r}}\!=\!16\!\times\!16$.
It can be seen from Fig. \ref{fig:channel_H_nmse} that the NMSE of the proposed NOMP method is always better than that of the OMP method, and the gap between the NOMP and OMP methods increases {with the SNR increase.}
Moreover, it can be checked that the performance of the OMP method is limited by the sampling rate of searching and the extended Newton method can further enhance the estimation performance.

\section{Conclusion}
In this letter, we have presented a wideband channel estimation scheme for an RIS assisted mmWave MIMO system.
We have firstly described the two individual channels and the frequency response of the received signal, where the wideband effect is considered.
Then, we {have expressed channel estimation as a parameter recovery problem} and {utilized} the NOMP algorithm to estimate the channel with a few OFDM pilot symbols.
Moreover, we have derived the CRLB for channel estimates.
{Simulation results have verified that the proposed channel estimation scheme achieves promising NMSE performances, especially at low SNR and with a few pilot symbols.}

\appendix
From \eqref{eq:F_pa_pa}, we can derive $\boldsymbol{\mathcal{F}}(\boldsymbol{\xi}_{p},\boldsymbol{\xi}_{p})$ as
\begin{align}
\boldsymbol{\mathcal{F}}(\boldsymbol{\xi}_{p},\boldsymbol{\xi}_{p})
=&\frac{2}{\sigma_{\text{v}}^2}\!\!\left[\begin{array}{ccccc}
\!\!\!\!1\!\!\! & \!\!\!|g_{p}| \!\!\!&\!\!\! |g_{p}|\!\!\!&\!\!\! |g_{p}|\!\!\!&\!\!\! 2\pi \Delta f|g_{p}|\!\!\!\! \\
\!\!\!\!|g_{p}|\!\!\!&\!\!\!  |g_{p}|^{2} \!\!\!&\!\!\!  |g_{p}|^{2} \!\!\!&\!\!\!  |g_{p}|^2 \!\!\!&\!\!\! 2\pi\Delta f|g_{p}|^{2}\!\!\!\! \\
\!\!\!\!|g_{p}|\!\!\!&\!\!\!  |g_{p}|^{2} \!\!\!&\!\!\!  |g_{p}|^{2} \!\!\!&\!\!\! |g_{p}|^2 \!\!\!&\!\!\!  2\pi\Delta f|g_{p}|^{2}\!\!\!\! \\
\!\!\!\!|g_{p}| \!\!\!&\!\!\!  |g_{p}|^{2} \!\!\!&\!\!\!  |g_{p}|^{2}\!\!\!&\!\!\!  |g_{p}|^2 \!\!\!&\!\!\! 2\pi\Delta f|g_{p}|^{2}\!\!\!\!\\
\!\!\!\!2\pi\Delta f|g_{p}|\!\!\!&\!\!\!  2\pi\Delta f|g_{p}|^{2} \!\!\!&\!\!\! 2\pi\Delta f|g_{p}|^{2} \!\!\!&\!\!\!  2\pi\Delta f|g_{p}|^{2} \!\!\!&\!\!\! (2\pi\Delta f|g_{p}|)^{2}\!\!\!\!
\end{array}\right]\notag\\
&\odot\!\!\!\left[\begin{array}{ccccc}
\!\!\!\!\!\Re\{[\mathbf{B}_{p}]\!_{1\!,\!1}\}\!\! \!&\! \!\! -\Im\{[\mathbf{B}_{p}]\!_{1\!,\!1}\}\!\! \!&\!\! \! \Re\{[\mathbf{B}_{p}]\!_{1\!,\!2}\}\! \!\!&\!\!  \!\Re\{[\mathbf{B}_{p}]\!_{1\!,\!3}\} \! \!\!&\!\!  \!\Im\{[\widetilde{\mathbf{B}}_{p}]\!_{1\!,\!1}\}\!\!\!\! \!\!\\
\!\!\!\!\!\Im\{[\mathbf{B}_{p}]\!_{1\!,\!1}\}\! \!\!\!&\!\! \! \!\Re\{[\mathbf{B}_{p}]\!_{1\!,\!1}\}\!\! \!\!&\!\! \! \! \Im\{[\mathbf{B}_{p}]\!_{1\!,\!2}\}\!\! \!\!&\!\!  \!\!\Im\{[\mathbf{B}_{p}]\!_{1\!,\!3}\}\!\! \!\!&\!\! \! \! -\Re\{[\widetilde{\mathbf{B}}_{p}]\!_{1\!,\!1}\}\!\! \!\!\!\!\\
\!\!\!\!\!\Re\{[\mathbf{B}_{p}]\!_{2\!,\!1}\}\!\! \!\!&\!\!  \!\!-\Im\{[\mathbf{B}_{p}]\!_{2\!,\!1}\}\! \!\!\!& \!\!\!\! \Re\{[\mathbf{B}_{p}]\!_{2\!,\!2}\}\!\! \!\!&\!\!  \!\!\Re\{[\mathbf{B}_{p}]\!_{2\!,\!3}\} \!\! \!\!&\!\!\!  \!\Im\{[\widetilde{\mathbf{B}}_{p}]\!_{2\!,\!1}\}\!\! \!\!\!\!\\
\!\!\!\!\!\Re\{[\mathbf{B}_{p}]\!_{3\!,\!1}\}\!\! \!\!&\!\!  \!\!-\Im\{[\mathbf{B}_{p}]\!_{3\!,\!1}\}\!\! \!\!&\!\!\!  \! \Re\{[\mathbf{B}_{p}]\!_{3\!,\!2}\}\!\! \!\!&\!\!  \!\!\Re\{[\mathbf{B}_{p}]\!_{3\!,\!3}\}\! \!\!\!&\!\! \! \! \Im\{[\widetilde{\mathbf{B}}_{p}]\!_{3\!,\!1}\}\!\!\!\!\!\!\\
\!\!\!\!\!-\Re\{[\widetilde{\mathbf{B}}_{p}]\!_{1\!,\!1}\}\! \!\!&\! \! \!-\Re\{[\widetilde{\mathbf{B}}_{p}]\!_{1\!,\!1}\}\!\! \!&\!  \! \!-\Im\{[\widetilde{\mathbf{B}}_{p}]\!_{1\!,\!2}\}\! \!\!&\!  \! \!-\Im\{[\widetilde{\mathbf{B}}_{p}]\!_{1\!,\!3}\}\! \!\!&\! \! \!\Re\{C_{p}\}\!\!\!\!\!\!
\end{array}\right]
\end{align}
where $\breve{\mathbf{a}}_{k,m,b}\triangleq[\mathbf{a}_{\text{B}}(k\Delta f,\theta_{\text{b}})]_b\mathbf{a}^T_{\text{R}}(k\Delta f,\phi_{\text{r}},\psi_{\text{r}}) \boldsymbol{\Omega}_m$, $\breve{A}_{k,m,b,p}\triangleq \breve{\mathbf{a}}_{k,m,b}\mathbf{a}_{\text{R}}(k\Delta f,\phi_p,\psi_p)$, $\mathbf{B}_{p}\triangleq\sum_{k,m,b}\mathbf{D}_{k,m,b,p}$,
$\widetilde{\mathbf{B}}_{p}\triangleq\sum_{k,m,b}k\mathbf{D}_{k,m,b,p}$,
$C_p \triangleq \sum_{k,m,b}k^2\breve{A}_{k,m,b,p}^{*}\breve{A}_{k,m,b,p}$,
\begin{align}
\mathbf{D}_{k,m,b,p}\!\!\triangleq\!\!\left[\begin{array}{ccc}
\!\!\breve{A}_{k,m,b,p}^{*}\!\! \\
\!\!\frac{\partial \breve{A}_{k,m,b,p}^*}{\partial \phi_p}\!\!\\
\!\!\frac{\partial \breve{A}_{k,m,b,p}^*}{\partial \psi_p}\!\!
\end{array}\right]\!\!\left[\begin{array}{ccc}
\!\!\breve{A}_{k,m,b,p} \!\!&\!\!
\frac{\partial \breve{A}_{k,m,b,p}}{\partial \phi_p}\!\!&\!\!
\frac{\partial \breve{A}_{k,m,b,p}}{\partial \psi_p}
\end{array}\!\!\right],
\end{align}
\begin{align}
&\left[\begin{array}{cccc}
\!\!\!\!\frac{\partial \breve{A}_{k\!,m\!,b\!,p}}{\partial \phi_p}\!\!\!\!\!\\
\!\!\!\!\frac{\partial \breve{A}_{k\!,m\!,b\!,p}}{\partial \psi_p}\!\!\!\!\!\\
\!\!\!\!\frac{\partial \breve{A}^*_{k\!,m\!,b\!,p}}{\partial \phi_p}\!\!\!\!\!\\
\!\!\!\!\frac{\partial \breve{A}^*_{k\!,m\!,b\!,p}}{\partial \psi_p}\!\!\!\!\!
\end{array}\right]
\!\!\!=\!\!\!\left[\begin{array}{cccc}
\!\!\!\!-\!\jmath2\pi(1\!\!+\!\!\frac{k\Delta f}{f_c})\breve{\mathbf{a}}_{k\!,m\!,b}\big({\mathbf{a}}_{\text{R}}(k\Delta f\!,\phi\!_p\!,\psi\!_p)\!\odot\!\dot{\boldsymbol{\varpi}}\!_{\phi\!_p}\big) \!\! \!\!\!\\
\!\!\!\!-\!\jmath2\pi(1\!\!+\!\!\frac{k\Delta f}{f_c})\breve{\mathbf{a}}_{k\!,m\!,b}\big({\mathbf{a}}_{\text{R}}(k\Delta f\!,\phi\!_p\!,\psi\!_p)\!\odot\!\dot{\boldsymbol{\varpi}}\!_{\psi\!_p}\big)\! \!\!\! \!\\
\!\!\!\!\jmath2\pi(1\!\!+\!\!\frac{k\Delta f}{f_c})\breve{\mathbf{a}}_{k\!,m\!,b}^*\big({\mathbf{a}}^*_{\text{R}}(k\Delta f\!,\phi\!_p\!,\psi\!_p)\!\odot\!\dot{\boldsymbol{\varpi}}\!_{\phi\!_p}\big)\!\!\!\! \!\\
\!\!\!\!\jmath2\pi(1\!\!+\!\!\frac{k\Delta f}{f_c})\breve{\mathbf{a}}_{k\!,m\!,b}^*\big({\mathbf{a}}^*_{\text{R}}(k\Delta f\!,\phi\!_p\!,\psi\!_p)\!\odot\!\dot{\boldsymbol{\varpi}}\!_{\psi\!_p}\big)\!\!\!\!\!
\end{array}\right],
\end{align}
\begin{align}
\dot{\boldsymbol{\varpi}}\!_{\phi\!_p}\!\!\!=\!\!\Big[\!0,\!\cdots\!,\!\frac{d\!\big(\!(N\!_{\text{x}}\!\!-\!\!1)\!\cos(\!\phi\!_p\!)\!\sin(\!\psi\!_p\!)\!\!+\!\!(N\!_{\text{y}}\!\!-\!\!1)\!\cos(\!\phi\!_p\!)\!\cos(\!\psi\!_p\!)\!\big)}{\lambda_{\text{c}}}\!\Big]^T,
\end{align}
\begin{align}
\dot{\boldsymbol{\varpi}}\!_{\psi\!_p}\!\!\!=\!\!\Big[\!0,\!\cdots\!,\!\frac{d\big(\!(N\!_{\text{x}}\!\!-\!\!1)\!\sin(\!\phi\!_p\!)\!\cos(\!\psi\!_p\!)\!\!-\!\!(N\!_{\text{y}}\!\!-\!\!1)\!\sin(\!\phi\!_p\!)\!\sin(\!\psi\!_p\!)\!\big)}{\lambda_{\text{c}}}\!\Big]^T.
\end{align}

\balance

%\begin{thebibliography}{1}
%
%	\bibitem{metis6wireless}
%	M.~METIS, ``wireless communications enablers for the twentytwenty information
%	society, eu 7th framework programme project,'' ICT-317669-METIS, Tech. Rep.
%	\end{thebibliography}

%\balance
%\bibliographystyle{IEEEtran}
%\bibliography{./bibtex/IEEEabrv,./bibtex/ref}

%\balance
%\bibliographystyle{IEEEtran}
%\bibliography{./bibtex/IEEEabrv,./bibtex/ref}
\end{document}